\documentclass[12pt]{article}
\usepackage{pstricks,pst-node,graphics,hyperref}
\begin{document}
\newcommand{\Z}{{\bf Z}}

\title{Discrete analogue computing with rotor-routers}
\author{James Propp}
\date{\today}

\maketitle

\noindent
{\sc Abstract}:
Rotor-routing is a procedure for routing tokens through a network
that can implement certain kinds of computation.
These computations are inherently asynchronous
(the order in which tokens are routed makes no difference)
and distributed (information is spread throughout the system).
It is also possible to efficiently check
that a computation has been carried out correctly
in less time than the computation itself required,
provided one has a certificate
that can itself be computed by the rotor-router network.
Rotor-router networks can be viewed as both
discrete analogues of continuous linear systems
and deterministic analogues of stochastic processes.

\begin{quotation}
Rotor-router networks are discrete analogues of
continuous linear systems such as electrical circuits;
they are also deterministic analogues of
stochastic systems such as random walk processes.
These analogies permit one to design rotor-router networks
to compute numerical quantities
associated with some linear and/or stochastic systems.
These distributed computations can behave stably 
even in the presence of significant disruption.
\end{quotation}

\section{\label{sec:intro} Introduction}

Rotor-routing is a protocol for routing tokens through a network,
where a network is represented as a directed graph
consisting of vertices and arcs.
In the simplest case,
where a vertex $v$ has two outgoing arcs $a_1$ and $a_2$,
the rotor-routing protocol dictates that
a token that leaves $v$ should leave along arc $a_1$
if the preceding token that left $v$ 
(which might be the same token at an earlier time or might not)
went along arc $a_2$, and vice versa. 
(See section~\ref{sec:three} for a discussion
of $n$-state rotor-routers for general values of $n$.)
The ``input'' to the computation is the
choice of rotors and the pattern of interconnection between them;
the output is a quantity associated with the evolution of the network
that can be measured by an observer watching the system
or stored in an output register by the network itself.

Rotor-router networks are more like
classical analog computers than like modern digital computers.
``Programming'' an analog computer means connecting the components,
and the ``output'' is the behavior of the system,
which one can measure in different numerical ways.
Classical analogue computing is possible
because different physical systems
can obey the same mathematical evolution laws;
if one can devise an electrical circuit to satisfy
the mathematical evolution laws one wishes to study,
the behavior of the electrical circuit
will faithfully mimic the behavior of
the actual system one wishes to study (a neuron, perhaps).
We show here that suitably constructed rotor-router networks
display similar fidelity to two sorts of (very simple) systems:
discrete random network flows, 
discussed in section~\ref{sec:random}, 
and continuous deterministic network flows, 
discussed in section~\ref{sec:divisible}.

Classical analogue computing is successful within its domain of applicability
because (a) the wealth of available components
permits one to embody a wide variety of evolution laws,
(b) a single constructed circuit can be driven in many ways,
and (c) a circuit being driven in a particular way
can be measured in a wide variety of ways;
(b) and (c) taken together offer the experimenter
a very rich picture of response characteristics of the system.
For discrete deterministic network flow models
(such as rotor-routing or, more generally,
abelian distributed processes, as described in Dhar (1999)), 
we have only a limited stockpile of components,
and it is unclear what class of models they can simulate.
(For instance, we do not know how to use models of this kind
to simulate linear systems with impedance as well as resistance.)
The rotor-router systems described in this article
also admit no driving terms or other form of ``input'',
other than the choice of how many tokens to feed into the network
(which determines the fidelity of the simulation:
the more tokens one feeds into the system,
the higher its fidelity to the system being simulated).
As a small consolation,
one can take different sorts of measurements
of a single rotor-router network
to determine different numerical characteristics
of the model it is simulating
(e.g., the respective current flow along different edges
in a circuit of resistors).
But, as one early reader of this article wondered,
if all that rotor-router networks can do
is simulate simple systems like networks of resistors
(or more generally solve Dirichlet problems on graphs),
of what use are they?

Our answer is that, although the computational powers 
of the networks described here are rather weak,
they can be viewed as prototypes of a style of computation
that might, with a suitably enlarged toolkit, 
lead to more interesting applications.

Specifically,
within their (currently very narrow) domain of applicability,
networks that implement rotor-routing
can carry out parallel computations with four noteworthy features
(the first two holding generally, 
and the last two holding under certain circumstances):

\begin{itemize}
\item Asynchronous: the order in which steps occur does not affect
the outcome of the computation.
\item Distributed: information is stored throughout the network.
\item Robust: even if errors occur (e.g., some tokens are routed along 
the wrong arc), the outcome of the computation will not be greatly affected.
More specifically, the error in the answer grows merely linearly in the
error rate.
\item Verifiable: each computation can be used to create a certificate
that can later be used to verify the outcome of the computation
in less time than the computation itself required.  
This is a consequence of the fact that
the evolution of the system satisfies a least action principle.
\end{itemize}

One reason for the tractable nature of rotor-router systems
is that, although a rotor-router system is nonlinear,
it can be viewed as an approximation to a continuous linear model.
This linear model in turn can be construed as the average-case behavior
of a discrete random network-propagation model.
This is not coincidental, as rotor-routing was invented circa 2000
by this author as a way of derandomizing such random systems
while retaining their average-case behavior.
A key technical tool in the analysis of rotor-router systems
is the existence of dynamical invariants
obtained by simply adding together
many locally-defined quantities;
in particular, these invariants are used
to prove the robustness property stated above,
and indeed to prove that the long-term behavior of a rotor-router system
mimics the behavior of both discrete stochastic network flow
and continuous deterministic network flow.

\section{\label{sec:three} Three network flow models}

\subsection{\label{sec:rotor} Rotor-routing}

An $n$-state rotor-router at a vertex $v$
has $n$ states (numbered 1 through $n$)
and its $i$th state is associated with an arc $a_i$
pointing from $v$ to a neighboring vertex.
We denote the arc from $v$ to $w$ by $(v,w)$.
When $v$ receives a token (which we will hereafter call a ``chip''
for historical reasons)
the state of the rotor at $v$ is incremented by 1
(unless the state was $n$, in which case it becomes 1),
and the chip is sent along the arc
associated with the new state of the rotor.
That is, if $v$ receives a chip when its rotor is in state $i$,
the rotor advances to state $i+1$ and sends the chip along arc $a_{i+1}$
(where $n+1$ is taken to be 1).
It is permitted to have $a_i = a_j$ with $i \neq j$.
We let $p(v,w) = \#\{1 \leq i \leq n: a_i = (v,w)\}/n$,
i.e., the proportion of rotor-states at $v$ pointing to $w$,
so that $\sum_w p(v,w) = 1$ for all $v$.

\begin{figure}
\centering
\begin{pspicture}(0,0)(8.0,2.0)
\rput(1.8,1.8){\circlenode{A}{\ \ }}
\rput(4.2,1.8){\circlenode{B}{\ \ }}
\rput(5.4,1.8){\circlenode{C}{\ \ }}
\rput(6.6,1.8){\circlenode{D}{\ \ }}
\rput(7.8,1.8){\rnode{E}{In}}
\rput(0.4,0.5){\rnode{F}{Out2}}
\rput(1.8,0.5){\rnode{G}{Out1}}
\rput(4.2,0.5){\rnode{H}{Out1}}
\rput(5.4,0.5){\rnode{I}{Out1}}
\rput(6.6,0.5){\rnode{J}{Out1}}
\rput(0.4,1.8){\rnode{K}{\ }}
\rput(2.8,1.8){\rnode{L}{\ . . .\ }}
\ncline[arrowsize=5pt]{->}{E}{D}
\ncline[arrowsize=5pt]{->}{D}{C}\nbput{0}
\ncline[arrowsize=5pt]{->}{C}{B}\nbput{0}
\ncline[arrowsize=5pt]{->}{B}{L}\nbput{0}
\ncline[arrowsize=5pt]{->}{L}{A}
\ncline[arrowsize=5pt]{->}{A}{G}\naput{1}
\ncline[arrowsize=5pt]{->}{B}{H}\naput{1}
\ncline[arrowsize=5pt]{->}{C}{I}\naput{1}
\ncline[arrowsize=5pt]{->}{D}{J}\naput{1}
\ncarc[arrowsize=5pt,arcangle=-45]{->}{A}{F}\nbput{0}
\end{pspicture}
\caption{\label{fig:counter} A binary counter made of rotor-routers.}
\end{figure}
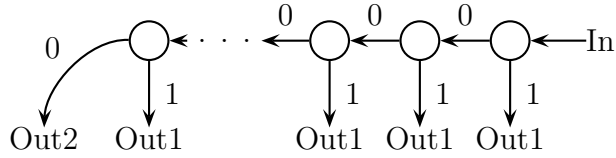

Figure~\ref{fig:counter} shows a binary counter 
(aka unary-to-binary converter)
consisting of a chain of $m$ rotor-routers. 
It should be viewed as an open system that can be connected
to other rotor-router systems to form a larger network,
with chips being fed into it along an input line
and exiting from it along two output lines. 
Each rotor-router in the chain except the ones at the ends
receives chips from the rotor-router to its right
and sends chips to the rotor-router to its left
and to the first output line.
The rotor-router at the far right
receives chips only from the input line,
and the rotor-router at the far left
sends chips to both the first and second output lines.
For this particular network,
it is more convenient to number the states 0 and 1.
When a rotor-router in state 0 receives a chip,
it changes its state to 1
and sends the chip along the first output line;
when a rotor-router in state 1 receives a chip,
it changes its state to 0
and sends the chip to the next rotor-router to its left.
If all rotors were initially in state 0,
then after $N<2^m$ chips have passed through the binary counter,
the states of the rotors, read from left to right,
will be the base two representation of the integer $N$.
When the $2^m$th chip is added,
it will cause all the rotors to return to state 0
and send the chip along the second output line,
indicating that an overflow has occurred.
Inasmuch as two-state rotor-routers are little more
than flip-flops,
it is not surprising that they can be used in this way
to carry out binary addition.
(The network of Figure 1 only implements addition of 1,
but with more input lines it can do addition of $m$-bit binary numbers.)

\begin{figure}
\centering
\begin{pspicture}(2.5,1.5)(9.5,7.5)
\rput(4,2){\circlenode{A}{\ \ }}
\rput(6,2){\circlenode{B}{\ \ }}
\rput(8,2){\circlenode{C}{\ \ }}
\rput(4,4){\circlenode{D}{\ \ }}
\rput(6,4){\circlenode{E}{\ \ }}
\rput(8,4){\circlenode{F}{\ \ }}
\rput(4,6){\circlenode{G}{\ \ }}
\rput(6,6){\circlenode{H}{\ \ }}
\rput(3.0,6){\rnode{I}{In}}
\rput(5.0,7){\rnode{J}{Out1}}
\rput(7.0,7){\rnode{K}{Out2}}
\rput(3.0,5.0){\rnode{L}{Out1}}
\rput(9.0,5.0){\rnode{M}{Out2}}
\ncarc[arcangle=-10,arrowsize=5pt]{->}{A}{B}
\ncarc[arcangle=-10,arrowsize=5pt]{->}{B}{A}
\ncarc[arcangle=-10,arrowsize=5pt]{->}{B}{C}
\ncarc[arcangle=-10,arrowsize=5pt]{->}{C}{B}
\ncarc[arcangle=-10,arrowsize=5pt]{->}{D}{E}
\ncarc[arcangle=-10,arrowsize=5pt]{->}{E}{D}
\ncarc[arcangle=-10,arrowsize=5pt]{->}{E}{F}
\ncarc[arcangle=-10,arrowsize=5pt]{->}{F}{E}
\ncarc[arcangle=-10,arrowsize=5pt]{->}{G}{H}
\ncarc[arcangle=10,arrowsize=5pt]{->}{A}{D}
\ncarc[arcangle=10,arrowsize=5pt]{->}{D}{A}
\ncarc[arcangle=10,arrowsize=5pt]{->}{B}{E}
\ncarc[arcangle=10,arrowsize=5pt]{->}{E}{B}
\ncarc[arcangle=10,arrowsize=5pt]{->}{C}{F}
\ncarc[arcangle=10,arrowsize=5pt]{->}{F}{C}
\ncarc[arcangle=10,arrowsize=5pt]{->}{E}{H}
\ncarc[arcangle=10,arrowsize=5pt]{->}{H}{E}
\ncarc[arcangle=10,arrowsize=5pt]{->}{G}{D}
\ncarc[arcangle=-10,arrowsize=5pt]{->}{D}{L}
\ncarc[arcangle=10,arrowsize=5pt]{->}{H}{J}
\ncarc[arcangle=-10,arrowsize=5pt]{->}{H}{K}
\ncarc[arcangle=10,arrowsize=5pt]{->}{F}{M}
\ncline[arrowsize=5pt]{->}{I}{G}
\end{pspicture}
\caption{\label{fig:electric} An electrical network simulator 
made of rotor-routers.}
\end{figure}
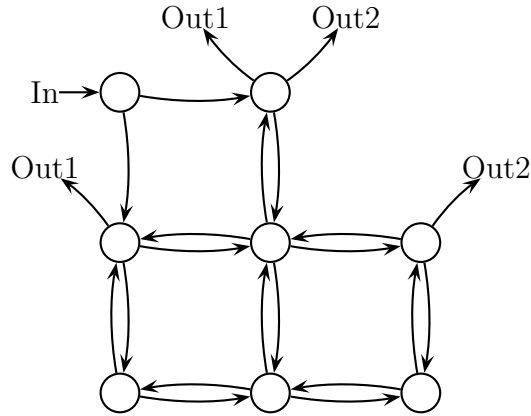

\begin{figure}
\centering
\begin{pspicture}(0.5,0.5)(3.5,3.5)
\psline(1,1)(1.2,1)(1.25,1.10)(1.35,0.90)(1.45,1.10)(1.55,0.90)(1.65,1.10)(1.75,0.90)(1.80,1)(2.0,1)
\psline(2,1)(2.2,1)(2.25,1.10)(2.35,0.90)(2.45,1.10)(2.55,0.90)(2.65,1.10)(2.75,0.90)(2.80,1)(3.0,1)
\psline(1,2)(1.2,2)(1.25,2.10)(1.35,1.90)(1.45,2.10)(1.55,1.90)(1.65,2.10)(1.75,1.90)(1.80,2)(2.0,2)
\psline(2,2)(2.2,2)(2.25,2.10)(2.35,1.90)(2.45,2.10)(2.55,1.90)(2.65,2.10)(2.75,1.90)(2.80,2)(3.0,2)
\psline(1,3)(1.2,3)(1.25,3.10)(1.35,2.90)(1.45,3.10)(1.55,2.90)(1.65,3.10)(1.75,2.90)(1.80,3)(2.0,3)
\psline(2,3)(2.2,3)(2.25,3.10)(2.35,2.90)(2.45,3.10)(2.55,2.90)(2.65,3.10)(2.75,2.90)(2.80,3)(3.0,3)
\psline(1,1)(1,1.2)(0.90,1.25)(1.10,1.35)(0.90,1.45)(1.10,1.55)(0.90,1.65)(1.10,1.75)(1,1.80)(1,2.0)
\psline(1,2)(1,2.2)(0.90,2.25)(1.10,2.35)(0.90,2.45)(1.10,2.55)(0.90,2.65)(1.10,2.75)(1,2.80)(1,3.0)
\psline(2,1)(2,1.2)(1.90,1.25)(2.10,1.35)(1.90,1.45)(2.10,1.55)(1.90,1.65)(2.10,1.75)(2,1.80)(2,2.0)
\psline(2,2)(2,2.2)(1.90,2.25)(2.10,2.35)(1.90,2.45)(2.10,2.55)(1.90,2.65)(2.10,2.75)(2,2.80)(2,3.0)
\psline(3,1)(3,1.2)(2.90,1.25)(3.10,1.35)(2.90,1.45)(3.10,1.55)(2.90,1.65)(3.10,1.75)(3,1.80)(3,2.0)
\psline(3,2)(3,2.2)(2.90,2.25)(3.10,2.35)(2.90,2.45)(3.10,2.55)(2.90,2.65)(3.10,2.75)(3,2.80)(3,3.0)
\dotnode(1,3){A}
\dotnode(3,3){B}
\rput(0.8,3.2){\it a}
\rput(3.2,3.2){\it b}
\end{pspicture}
\caption{\label{fig:circuit} The electrical network being simulated
in Figure~\ref{fig:electric}.  Nodes $a$ and $b$ correspond to
lines Out1 and Out2, respectively.}
\end{figure}
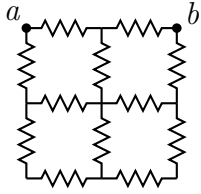

Figure~\ref{fig:electric} shows a seemingly very different way
of using rotor-routers as computational elements.
The network here is computing the effective conductance
of the 3-by-3 square grid of unit resistors shown in Figure~\ref{fig:circuit},
as measured between corners $a$ and $b$.
The reader should imagine that the outgoing arcs
from each of the eight vertices
are numbered counterclockwise $1$ through $n$,
where $n$ is the number of outgoing arcs from the vertex.
These labels have been omitted from Figure~\ref{fig:electric};
for our purposes,
what matters is the cyclic ordering of the arcs,
not their precise numbering.
(Indeed, there is nothing particularly special
about the counterclockwise ordering
of the arcs emanating from a vertex;
the results of this paper
remain qualitatively true for arbitrary orderings,
though the quantitative results
depend on which ordering is chosen.)
After $N$ chips have entered the network through the input line
and exited through one of the two output lines,
the number that left through the second output line, times two,
divided by the total number of chips that have gone through the network,
is approximately equal to the effective conductance of the network,
and the discrepancy between the two quantities
goes to zero at rate constant/$N$ as $N$ gets large.
(This will be explained and generalized in subsection~\ref{sec:invariant}.)
One can view the chips entering the network along the input line
as constituting a unary representation 
of the desired accuracy of the simulation.

\subsection{\label{sec:random} Random routing}

It is instructive to consider a variant of rotor-routing
in which each successive chip is routed along a random arc
(rather than the next arc in the pre-specified rotation sequence).
Then each chip is simply executing a random walk,
where the probability that a walker at $v$ will take a step to $w$
is $p(v,w)$.  As is well known 
(Doyle and Snell, 1984; Lyons and Peres, 2010), 
there is an intimate connection between
random walks on finite (undirected) graphs and electrical networks.
Indeed, the effective conductance $C_{\rm eff}$ of a resistive network
as measured between two vertices $a$ and $b$
satisfies the formula $C_{\rm eff} / c_a = p_{\rm esc}$
where the local conductance $c_a$ is the sum of the conductances
of the edges joining $a$ to the rest of the network
and the escape probability $p_{\rm esc}$ is the probability
that a random walker starting from $a$ 
will reach $b$ before returning to $a$.
So, if one lets $N$ walkers walk randomly in the graph
shown in Figure~\ref{fig:electric}
(or, equivalently, if one routes $N$ chips through the directed graph
shown in Figure~\ref{fig:circuit}
using random routing),
and if one removes the walkers when they arrive at $a$ or $b$,
then the number of walkers that exit at $b$, times two, divided by $N$, 
will converge to the effective conductance of the electrical network
between $a$ and $b$. 
However, the discrepancy will be on the order of constant/$\sqrt{N}$.
Rotor-routing brings the order of the discrepancy down to constant/$N$.

\subsection{\label{sec:divisible} Divisible routing}

Yet another variant of rotor-routing that is worth considering
is {\it divisible routing}.
In this scenario, chips may be subdivided,
and our rule is that when a chip of any size
is received at an $n$-state vertex,
it is split into $n$ smaller equal-sized chips.
It is helpful here to change one's language
and speak of fluid flowing through the system,
where the fluid at a vertex
gets divided equally among the outgoing arcs.
Both random routing and rotor-routing are
discrete approximations to the continuous divisible routing model.
This model is linear,
and one reason for the tractability of both
the random routing and rotor-rounding models
is that they can be seen as variations of the linear model. 
(In fact, the amount of fluid that leaves the network at $b$
in the divisible-routing case is exactly equal to
the expected number of chips that leave the network at $b$
in the random routing case.)

It is fairly obvious that for the random routing model,
instead of imagining the chips as passing through the system sequentially
we could imagine them as passing through the system simultaneously;
as long as they do not interact,
and they individually behave randomly,
the (random) number of them that exit the network
at one vertex versus another
should follow the same probability law
as in the sequential case.
It is less obvious, but nonetheless true
(and not too hard to prove; see Holroyd {\it et al.}, 2008), 
that a similar property holds for rotor-routing;
this is called the {\it abelian property\/} of rotor-routing.
Specifically, imagine that we have a number of indistinguishable chips
located at the vertices of a rotor-router network.
To avoid degeneracy, suppose that the network is connected,
and that for each vertex $v$
there is a path of vertices $v_0, v_1, v_2, \dots, v_m$ such that $v_0=v$, 
for all $i<m$ there is a rotor-state at $v_i$ that points to $v_{i+1}$,
and $v_m$ has a state that causes a chip to leave the network.
Then for any asynchronous routing of the chips,
all the chips must eventually exit the network along an output line,
and the number of chips that exit along any particular outline line
is independent of the order in which routing-moves occur.

Here our notion of asynchronous routing is that at each instant,
at most one chip makes a move,
and it does so by advancing the state of the rotor
at the site it currently occupies
and then moving to the site that the new rotor-state points to.

If one wants to permit several chips
to move at the same time,
the model can accommodate this,
provided one makes sure that this does not cause a ``jam''
but rather causes two colliding chips
that want to arrive at $v$ simultaneously
to form a queue.
Note that since the chips are assumed to be indistinguishable,
forming such a queue does not require making a random choice
or breaking the symmetry of the system in any way.

Note also that there is no difference between, on the one hand,
sending $N$ indistinguishable chips into the network 
along the input line and seeing which output lines they take,
and, on the other hand,
sending a single chip through the network $N$ times
and seeing which output lines it takes,
where it is returned to the network along the input line
each time it exits along an output line.
In both situations, the only quantity that we attend to
is the number of times each respective output line is used;
these counts must add up to $N$, 
and the abelian property assures us that the count associated
with a particular output line is the same under the two scenarios.

\section{\label{sec:simulation} Simulation with rotor-routers}

\subsection{\label{sec:resist} A general theorem for resistive networks}

Suppose we have a finite network of resistors
such that the conductance $c_{v,w}$ between any two nodes $v,w \in V$
is a rational number.
Let $C_{\rm eff}$ denote the effective conductance of this network
as measured between chosen nodes $a$ and $b$
(this is the amount of current that would flow from $a$ to $b$
if we attached these two nodes to a 1-volt battery,
clamping the voltage at $a$ at 0 and the voltage at $b$ at 1).
As a technicality,
we need to modify the graph
by introducing two copies of the vertex $a$,
which we will call $a$ and $a'$.
(Looking ahead to the random walk and rotor walk models,
chips enter the network at $a$
and exit from $a'$ or $b$;
it is necessary to distinguish between $a$ and $a'$
since a walker that is at $a$ for the first time
can continue walking within the network
but a walker that is at $a$ for the second time
must immediately exit.)
We define the local conductance at $v$ as $c_v = \sum_w c_{v,w}$,
the sum of the conductances of the edges incident with $v$.
Consider a rotor-router network with two vertices $a,a'$
corresponding to the node $a$ of the original network
and with a single vertex corresponding to every other node of the network
(including $b$),
such that for all $v \neq a'$ and $w \neq a$,
$p(v,w)$ (the proportion of the rotor-states at $v$ that point to $w$)
is equal to the normalized conductance $c_{v,w}/c_v$.
We have a first output line from $a'$
and a second output line from $b$.
After $N$ chips have entered and left the network,
the number that left through $b$, times $c_a$,
divided by the total number of chips that have gone through the network,
is approximately equal to the effective conductance of the network,
and the discrepancy between the two quantities
is bounded by an explicit (network-dependent) constant 
times $1/N$ for all $N$.

The proof does not appear in the literature,
but it is easily obtained by combining results in
Doyle and Snell (1984) 
with results in Holroyd and Propp (2010).  
The former provides the link 
between (purely resistive) electrical networks and random walks,
and the latter provides the link between random walks and rotor walks.

Note that since vertices $a'$ and $b$ each have only a single outward arc,
these vertices are dispensable;
we can replace every arc to $a'$
by an arc that goes directly to the first output line
and every arc to $b$
by an arc that goes directly to the second output line.
This is how the rotor-router network shown in Figure~\ref{fig:electric}
was derived from the resistor network shown in Figure~\ref{fig:circuit}.
Lines In, Out1, and Out2 correspond 
to vertices $a$, $a'$, and $b$ respectively.

\subsection{\label{sec:invariant} Dynamical invariants}

A key tool in the proofs given by Holroyd and Propp (2010) 
is the existence of simple numerical dynamical invariants of rotor-routing.
These invariants are associated with the states of the rotors
and the locations of chips currently in the network.
A chips-and-rotors configuration
consists of an arrangement of chips on the vertices
as well as some configuration of the rotors
(i.e.\ some assignment of states to the respective rotors).
Since chips are indistinguishable, the arrangement of chips
is given by a non-negative function on the vertex set of the graph
that indicates the number of chips present at each vertex.
The value of a chips-and-rotors configuration
is equal to the sum of the values of all the chips
and the values of all the rotors,
where the value of a chip depends only on its location
and the value of a rotor depends only on its state.
If we have chosen our value-function with care,
then the operation of changing the state of a rotor
and the operation of moving a chip in accordance
with the new state of the rotor
perfectly offset one another,
resulting in no net change in the value of the configuration.

Such numerical invariants exist for both
the network shown in Figure~\ref{fig:counter}
and the network shown in Figure~\ref{fig:electric},
and indeed serve as a unifying link between the two sorts of networks,
so we will discuss the two examples in turn
before turning to the more general situation.

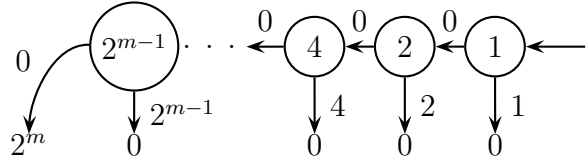
\begin{figure}
\centering
\begin{pspicture}(0,0)(8.8,2.0)
\rput(1.8,1.8){\circlenode{A}{$2^{m-1}$}}
\rput(4.2,1.8){\circlenode{B}{\ 4\ }}
\rput(5.4,1.8){\circlenode{C}{\ 2\ }}
\rput(6.6,1.8){\circlenode{D}{\ 1\ }}
\rput(7.8,1.8){\rnode{E}{\ \ }}
\rput(0.4,0.5){\rnode{F}{$2^m$}}
\rput(1.8,0.5){\rnode{G}{0}}
\rput(4.2,0.5){\rnode{H}{0}}
\rput(5.4,0.5){\rnode{I}{0}}
\rput(6.6,0.5){\rnode{J}{0}}
\rput(0.4,1.8){\rnode{K}{\ }}
\rput(2.8,1.8){\rnode{L}{\ . . .\ }}
\ncline[arrowsize=5pt]{->}{E}{D}
\ncline[arrowsize=5pt]{->}{D}{C}\nbput{0}
\ncline[arrowsize=5pt]{->}{C}{B}\nbput{0}
\ncline[arrowsize=5pt]{->}{B}{L}\nbput{0}
\ncline[arrowsize=5pt]{->}{A}{G}\naput{$2^{m-1}$}
\ncline[arrowsize=5pt]{->}{B}{H}\naput{4}
\ncline[arrowsize=5pt]{->}{C}{I}\naput{2}
\ncline[arrowsize=5pt]{->}{D}{J}\naput{1}
\ncarc[arrowsize=5pt,arcangle=-45]{->}{A}{F}\nbput{0}
\end{pspicture}
\caption{\label{fig:countervalues} The values of vertices and arcs 
for Figure~\ref{fig:counter}.}
\end{figure}

For the example shown in Figure~\ref{fig:counter}, 
label the sites as 0, 1, 2, etc.\ from right to left.
A chip at vertex $i$ has value $2^i$,
and the rotor at vertex $i$ has value $2^i$ 
when it points towards the first output line
and value 0 otherwise (i.e.\ when it points toward the left
or, in the case of the leftmost vertex,
when it points to the second output line).
It is easy to check that as a chip moves through the network,
changing rotor-states as it goes,
each operation of advancing the rotor at $i$
and moving a chip away from $i$ along an outgoing arc
has no net effect
on the value of the chips-and-rotors configuration
(except when the chip leaves the leftmost vertex along the second output line).
Thus, if the system has no chips at the vertices,
the operation of adding a chip along the input line
and letting it propagate until it leaves the system
usually increases the value of the rotors by 1,
since adding the chip at the right
increases the value of the chips-and-rotors configuration by 1
and this value is unchanged as the chip
moves through the system and exits along the first output line.
The one exceptional case is when all rotors are in the state 1;
in this (``overflow'') case, 
adding a chip causes the rotors to revert to the all-0's state,
so the value of the system decreases by $2^m-1$,
where $m$ is the number of rotors in the chain.
See Figure~\ref{fig:countervalues}.
In this Figure, we have given the output lines
respective values 0 and $2^m$,
since with these values
the dynamical invariance property
holds even when overflow occurs.
When a chip enters the network and exits along the first output line,
the value of the rotors increases by 1;
when a chip enters the network and exits along the second output line,
the value of the rotors decreases by $2^m-1$.

For the example shown in Figure~\ref{fig:electric},
the value of a chip at a vertex $v$
is defined as the electrical potential of $v$
if in the corresponding electrical network
we clamp vertex $a'$ at voltage 0
and clamp vertex $b$ at voltage 1.
This electrical potential can also be interpreted as the probability
that a random walker starting from $v$
will arrive at $b$ before reaching $a'$.
Figure~\ref{fig:electricvalues} 
shows a way of assigning values to the rotor-states
so that the value of a chips-and-rotors configuration
is invariant under the combined operation
of updating the rotor at $v$
(rotating it counterclockwise to the next outgoing arc)
and sending a chip from $v$
to the neighbor that the rotor at $v$ now points to.
Here we give the first output line the value 0
and the second output line the value 1,
corresponding to the respective voltages at $a$ and $b$
in the original circuit,
and corresponding to the respective probabilities
that a random walker who starts at $a'$ or $b$
will leave the network at $b$.
Note that the value of $a$ is $.4$.
When a chip enters the network and exits along the first output line,
the value of the rotors increases by $.4-0=.4$
(the value of the input output line minus the value of the first output line);
when a chip enters the network and exits along the second output line,
the value of the rotors increases by $.4-1=-.6$
(the value of the input output line minus the value of the second output line),
i.e.\ decreases by $.6$.

\begin{figure}
\centering
\begin{pspicture}(2.5,1.5)(9.5,7.5)
\rput(4,2){\circlenode{A}{.4}}
\rput(6,2){\circlenode{B}{.5}}
\rput(8,2){\circlenode{C}{.6}}
\rput(4,4){\circlenode{D}{.3}}
\rput(6,4){\circlenode{E}{.5}}
\rput(8,4){\circlenode{F}{.7}}
\rput(4,6){\circlenode{G}{.4}}
\rput(6,6){\circlenode{H}{.5}}
\rput(5.0,7){\rnode{J}{0}}
\rput(7.0,7){\rnode{K}{1}}
\rput(3.0,5.0){\rnode{L}{0}}
\rput(9.0,5.0){\rnode{M}{1}}
\ncarc[arcangle=-10,arrowsize=5pt]{->}{A}{B}\nbput{.0}
\ncarc[arcangle=-10,arrowsize=5pt]{->}{B}{A}\nbput{.1}
\ncarc[arcangle=-10,arrowsize=5pt]{->}{B}{C}\nbput{.0}
\ncarc[arcangle=-10,arrowsize=5pt]{->}{C}{B}\nbput{.1}
\ncarc[arcangle=-10,arrowsize=5pt]{->}{D}{E}\nbput{.0}
\ncarc[arcangle=-10,arrowsize=5pt]{->}{E}{D}\nbput{.2}
\ncarc[arcangle=-10,arrowsize=5pt]{->}{E}{F}\nbput{.0}
\ncarc[arcangle=-10,arrowsize=5pt]{->}{F}{E}\nbput{.2}
\ncarc[arcangle=-10,arrowsize=5pt]{->}{G}{H}\nbput{.0}
\ncarc[arcangle=10,arrowsize=5pt]{->}{A}{D}\naput{.1}
\ncarc[arcangle=10,arrowsize=5pt]{->}{D}{A}\naput{.2}
\ncarc[arcangle=10,arrowsize=5pt]{->}{B}{E}\naput{.0}
\ncarc[arcangle=10,arrowsize=5pt]{->}{E}{B}\naput{.2}
\ncarc[arcangle=10,arrowsize=5pt]{->}{C}{F}\naput{.0}
\ncarc[arcangle=10,arrowsize=5pt]{->}{F}{C}\naput{.3}
\ncarc[arcangle=10,arrowsize=5pt]{->}{E}{H}\naput{.0}
\ncarc[arcangle=10,arrowsize=5pt]{->}{H}{E}\naput{.5}
\ncarc[arcangle=10,arrowsize=5pt]{->}{G}{D}\naput{.1}
\ncarc[arcangle=-10,arrowsize=5pt]{->}{D}{L}\naput{.3}
\ncarc[arcangle=10,arrowsize=5pt]{->}{H}{J}\nbput{.5}
\ncarc[arcangle=-10,arrowsize=5pt]{->}{H}{K}\naput{.0}
\ncarc[arcangle=10,arrowsize=5pt]{->}{F}{M}\nbput{.0}
\ncline[arrowsize=5pt]{->}{I}{G}
\end{pspicture}
\caption{\label{fig:electricvalues} The values of vertices and arcs 
for Figure~\ref{fig:electric}.}
\end{figure}

The situation for more general finite electrical networks is similar.
As in the specific example discussed above,
the value of a chip at a vertex $v$
is defined as the electrical potential of $v$
if in the corresponding electrical network
we clamp vertex $a'$ at voltage 0
and clamp vertex $b$ at voltage 1.
This is forced upon us by the abelian property:
Suppose that $v$ is neither $a'$ nor $b$,
and that we have an $n$-state rotor at $v$.
If we put $n$ chips at $v$,
then one possible way to evolve the system
is to have each of the $n$ chips take a single step,
causing the rotor at $v$ to undergo one full revolution.
Since the rotor configuration is now
exactly what it was before any of the $n$ chips moved,
we see that the total value of the $n$ chips
must be the same before and after.
That is, if we let $h(\cdot)$ denote the value of a chip
at a particular location,
we must have $n \: h(v) = \sum_w n\: p(v,w) \: h(w)$,
where $n \: p(v,w)$ is the number of rotor-states at $v$
that point to $w$.  That is, we must have 
$h(v) = \sum_w p(v,w) \: h(w) = \sum_w (c_{v,w}/c_v) \: h(w)$,
i.e., $c_v \: h(v) = \sum_w c_{v,w} \: h(w)$.
But Kirchhoff's voltage law
tells us that the voltage function has this property
(and indeed it is the only function with this property
satisfying the boundary conditions $h(a') = 0$, $h(b) = 1$).

As for the rotor-states, there is a way to assign values to the states
so that the total value of a chip-and-rotors configuration
is a dynamical invariant
(that is, it does not change as long as chips remain within the network).
First note that dynamical invariance holds if it holds ``locally'',
that is, if for all $v$ the value of a chip-and-rotor configuration
does not change when a chip moves from $v$ to another vertex.
So it suffices to focus on the vertices $v$ individually.
If we have an $n$-state rotor at $v$,
we can introduce $n$ unknowns for the values of the rotor-states.
Dynamical invariance at $v$ holds 
if the $n$ unknowns satisfy $n$ linear equations,
where the $i$th equation represents the condition
that the value of a chip at $v$
plus the value of the $i$th rotor-state at $v$ does not change
if the rotor at $v$ is advanced from state $i$ to state $i+1$
(where $n+1$ is interpreted as 1)
and the chip at $v$ moves from $v$ to the neighbor of $v$
associated with the $i+1$st rotor-state.
From the form of the equations
(each of which specifies the difference between two of the unknowns),
we see that the only problem that might arise
is that the sum of the equations might be inconsistent.
However, if we add the $n$ equations,
so that the values of the rotor-states drop out of the equation,
we are left with $n \: h(v) = \sum_w n\: p(v,w) \: h(w)$,
which we know is already satisfied by $h(\cdot)$.
Hence exists a one-parameter family of ways 
to assign values to the rotor-states at $v$
so that the operation of rotor-routing at $v$
preserves the sum of all chip-locations
and the value of the rotor-state at $v$.

One natural way to standardize the assignment of values to rotor-states
is to require that at each vertex $v$,
the state with smallest value has value 0.
Alternatively we could require that for every $v$
the average value of the rotor-states at $v$ is 0.
We have adopted the former approach for our examples.

Since one vertex is clamped at value 0
and one vertex is clamped at value 1,
every other vertex will have voltage between 0 and 1,
so that every chip has value between 0 and 1
regardless of its location.
It can be shown that the values of each rotor-state at $v$
can be chosen to lie between $0$ and $n_v/4$
if the rotor at $v$ is an $n_v$-state rotor.
This implies that every rotor configuration in the network
has value between 0 and $\frac14 \sum_v n_v$.

Recall that $p_{\rm esc}$ is the probability that
a walker that enters the electrical network at $a$
and does random walk with transition probabilities $p(v,w)$
reaches $b$ before returning to $a$.
When a chip enters the corresponding rotor network 
at $a$ (whose value is $p_{\rm esc}$)
and exits along the first output line (whose value is 0), 
the value of the rotors increases by $p_{\rm esc}-0 = p_{\rm esc}$;
when a chip enters the network at $a$
and exits along the second output line (whose value is 1),
the value of the rotors increases by $p_{\rm esc}-1$
(i.e.\ decreases by $1-p_{\rm esc}$).
Hence, if $N$ chips go through the system, 
with $N-K$ of them going to the first output line
and $K$ of them going to the second output line,
the net change in the value of the rotors will be an increase of 
$(N-K)(p_{\rm esc}) + (K)(p_{\rm esc}-1) = N(p_{\rm esc}) - K$.
However, the total value of the rotors
remains in some bounded interval
(the interval $[0,1.7]$ in the example 
shown in Figures~\ref{fig:electric} and~\ref{fig:electricvalues});
suppose this interval has width $c$
(where we showed above that $c \leq \frac14 \sum_v n_v$).
Then $|K - N p_{\rm esc}| \leq c$,
so that $|K/N - p_{\rm esc}| \leq c/N$.

This last inequality says that 
the number of chips that exited along the first output line,
divided by the total number of chips that have gone through the system,
differs from $p_{\rm esc}$ by at most a constant divided by $N$.

If we wished, we could combine the examples shown 
in Figures~\ref{fig:counter} and~\ref{fig:electric}
by having two binary counters of the kind shown in Figure~\ref{fig:counter},
one for each output line of the network shown in Figure~\ref{fig:electric},
serving as output registers.
That is, chips that left the electrical-network simulator
could be passed on to one binary counter or the other
(according to whether they left through $a'$ or $b$)
before leaving the system entirely.
Then, after $N$ chips had been fed into the compound system,
the two binary counters would record 
the number of exits along the respective output-lines,
which as remarked above would yield 
an approximation to the effective conductance
of the electrical network.
This observation underlines the fact that
the networks of Figures~\ref{fig:counter} and~\ref{fig:electric}
are really quite analogous:
both are doing arithmetic internally,
recording the system's current value
as a sum of many values residing at different vertices. 

Moreover, both networks can be construed
as deterministic analogues of random processes.
We have already seen that the second network
is an analogue of random walk
on the electrical circuit of Figure~\ref{fig:circuit}; 
likewise, the first network
is a derandomization of the random reset process
in which a counter (initially 0)
either increases by 1 or is reset to 0 at each step,
with each possibility occurring with probability 1/2,
except when the counter is $m-1$,
in which case the counter can only be reset to 0
at the next step.

Alternatively, one can view the network of Figure~\ref{fig:counter}
as a discrete analogue of the continuous flow process
that pushes one unit of fluid through an input line
and, at each junction,
sends half of the remaining fluid to the output line
and the remaining half on to the next junction.

Thus, a circuit that does binary counting
(as ``digital'' a process as one could imagine)
can be seen as a deterministic analogue of a stochastic system
or as a discrete analogue of a continuous linear system.

\subsection{\label{sec:onedim} An infinite one-dimensional Markov chain}

An example of using rotor-routers to compute
properties of an infinite Markov chain 
is described in Kleber (2005).  
We imagine a bug doing random walk on $\{-1,0,1,2,3,\dots\}$
so that at each time step it has probability 1/2 of going 1 to the right
and probability 1/2 of going 2 to the left,
where $-1$ and $0$ are absorbing states.
Elementary random walk arguments tell us
that the probability of the bug ending up in $\{-1,0\}$ is 1,
and that the probability of the bug ending up at $-1$ is 
$\tau = (-1+\sqrt{5})/2$.
If we tried this experiment with $N$ bugs doing random walk,
the expected number of bugs ending up at $-1$ would be $N \tau$,
with standard deviation $O(\sqrt{N})$.
In contrast, suppose we move chips through a rotor-router network
in which each location $i \geq 1$ has a 2-state rotor,
with one state that sends a chip to $i+1$
and one state that sends a chip to $i-2$.
Suppose moreover that the first time a chip leaves $i$ it is routed to $i-2$.
Then if we send $N$ chips through this system,
the number $K_N$ that leave via $-1$ 
differs from $N\tau$ by at most $\tau$, for all $N$.
That is, if one momentarily ignores the fact
that the system can be solved exactly
and tries to adopt a Monte Carlo approach to estimating the probability
that the bug arrives at $-1$ before it arrives at 0,
standard Monte Carlo has typical error $O(1/\sqrt{N})$
while derandomized Monte Carlo via rotor-routers has error $O(1/N)$.

(To convert the scenario of Kleber (2005)
into the scenario of this article,
create an input line that goes to 1
and output lines that lead from $-1$ and 0.)

\subsection{\label{sec:twodim} An infinite two-dimensional Markov chain}

Another example of computing with rotor-routers
is described in Holroyd and Propp (2010).  
Here the Markov chain being derandomized 
has state-space $\{(i,j): i,j \in \Z\}$
and we imagine a bug doing random walk 
so that at each time step it has probability 1/4 of going any of 
the four neighbors of the current site,
except that when the bug visits $(1,1)$,
it goes back to $(0,0)$.
The rule regarding $(1,1)$ may seem a bit strange,
but it was chosen to ensure
that the probability that a bug that starts at $\{(0,0)\}$
will eventually return to the set $\{(0,0), (1,1)\}$ is 1
and that the probability of the bug ending up at $(1,1)$ is $\pi/8$
(this is readily derived from the formula $a(1,1) = 4/\pi$
on page 149 of Spitzer, 1976). 
If we tried this experiment with $N$ bugs doing random walk,
the expected number of bugs ending up at $(1,1)$ would be $N \pi/8$,
with standard deviation $O(\sqrt{N})$.
On the other hand, if we move chips through a rotor-router network
in which each location $(a,b) \neq (1,1)$ has a 4-state rotor,
and if we send $N$ chips through this system,
the number $K_N$ that end up at $(1,1)$ differs from $N\pi/8$ by $O(\log N)$.
Indeed, the $\log N$ bound may be unduly pessimistic:
for all $N$ up through $10^4$
(the point up through which simulations have been conducted),
$K_N$ never differs from $N\pi/8$ by more than 2.1,
and for more than half the values of $N$ up through $10^4$,
$K_N$ differs from $N \pi / 8$ by less than 0.5
(that is, $K_N$ is actually the integer closest to $N \pi / 8$).

\section{\label{sec:properties} Properties of rotor-routing}

\subsection{\label{sec:async} Asynchronousness}

The propositions proved in Holroyd {\it et al.\/} (2008) 
establish an ``abelian property'' for the rotor-router model:
as long every chip that can be moved does eventually get moved,
the final disposition of the chips (that is,
the tallies of how many chips left the network along each arc)
does not depend on the order in which chips were moved.

\subsection{\label{sec:holo} Distributedness}

If we adopt the sequential point of view
and let a chip pass entirely through the network
before introducing a new chip
(or equivalently re-introducing the old chip) into the system,
so that there is never more than one chip in the system,
then we can ask, what sort of ``memory'' does the system possess
that enables it to compute quantities like the effective conductance?
As we have seen, the random router model can be used
to compute the effective conductance of a network of resistors,
albeit with error $O(1/\sqrt{N})$ rather than $O(1/N)$,
so there is a sense in which the information in the network
is stored in the pattern of connections
(since in the case of random routing,
nothing else is remembered by the system ---
of course there is also information in the location of the chip,
but the information content of the chip itself
is merely the logarithm of the number of vertices).

On the other hand, rotor-routing does better than random routing,
and we might say that the relevant information
is stored in the rotors.
Indeed, we can be more specific here,
and say that the information is stored in the values of the rotors, 
as defined above.
Recall the reason for the effectiveness of rotor-routing
as a way of estimating escape probabilities:
the number of escapes during the first $N$ rotor-walks,
minus $N$ times the escape probability,
is constrained to be equal to
the difference between the final value of the rotors
and the initial value of the rotors,
and this difference in turn is bounded in absolute value
by the difference between
the maximum possible value of the rotors
and the minimum possible value of the rotors.

\subsection{\label{sec:robust} Robustness}

For networks like the one shown in Figure~\ref{fig:counter},
robustness does not apply,
since some of the rotor-states have much larger values than others;
a mistake at a rotor whose states have a wide range of values
can have a great impact on the final answer.
(This is just a way of saying that a binary counter
can be very inaccurate if the high-order bits are changed.)

On the other hand, 
for networks like the one shown in Figure~\ref{fig:electric},
each vertex has potential between 0 and 1
and we may assume that the rotor at $v$ has states
that all take values between $0$ and $n_v/4$,
where the rotor at $v$ is an $n_v$-state rotor.
Suppose the network as a whole takes values
lying between $0$ and $c$. 
If a rotor were to advance to the wrong state,
this would affect the value of the system by only a small relative amount,
specifically, at most $d/4$,
where $d = \max_v n_v$ is the maximum number of outgoing arcs at any vertex.
Thus, in the notation of subsection~\ref{sec:sandpile},
we would have $|K - N p_{\rm esc} \pm d/4| \leq c$,
so that $|K/N - p_{\rm esc}| \leq c/N + d/4N$.
Likewise, if a site were to send a chip to the wrong neighbor
while advancing its rotor properly,
we could treat this mistake as if
the rotor had advanced improperly twice 
(once before and once after the incorrect routing),
so by the same reasoning
we can bound the inaccuracy of our estimate of $p_{\rm esc}$ by $c/N + d/2N$.
If many errors occur, say $\epsilon N$,
with rotors advancing improperly an $\epsilon$ proportion of the time,
the discrepancy between $C_{\rm eff}$
and the rotor-router network's approximation to this quantity
will be at most $c/N + \epsilon d / 2$.

Additionally, suppose that at some moment
(after a chip has left along an output line
and before it has returned to the network along an input line) 
we were to reset {\it all\/} the rotors in the system.
The chips-and-rotors value of the system would be reset
to some number between 0 and $c$,
and the performance bound $|K'/N' - p_{\rm esc}| \leq c/N'$ would apply,
where $N'$ is the number of chips that exited the system after the reset
and $K'$ is the number of those chips that exited along the first output line,
implying $|(K+K')/(N+N') - p_{\rm esc}| \leq 2c/(N+N')$.
In this sense, over-writing the states of the rotors
has only a small impact on the fidelity of the system,
as long as $c$ is small.
This is further support for the contention that
the system's most important form of memory is in the pattern of connections,
and that the function of the rotors is to enable the system
to make optimum use of those connections
to achieve as high fidelity as possible.

\subsection{\label{sec:verify} Verifiability}

The odometer function is defined as
the integer-valued function of the vertex-set
that records for each $v$
the number of times $v$ sent a chip to a neighbor.
Levine noticed that
the odometer functions satisfies a least action principle
that makes it fairly simple to check that
a proposed odometer function is valid,
relative to a specified initial configuration of the rotors.
(This extends an observation of Moore and Machta (2000) 
in the context of the sandpile model,
discussed below in subsection~\ref{sec:sandpile},
as well as Deepak Dhar's observation that
the sandpile model satisfies what he called
the ``lazy man's least action principle''.)
The number of operations required to check a proposed odometer function
is on the order of the number of edges
times the logarithm of the maximum value of the odometer function.

Friedrich and Levine (2010) 
made use of the least action principle
in their study of two-dimensional rotor-router aggregation
(discussed below in subsection~\ref{sec:grow}).
In particular, their way of building the $N$-particle aggregate
appears to have running-time $\Theta(N \log N)$ rather than $\Theta(N^2)$
(the latter being the amount of time required
to carry out a straightforward simulation).
This has enabled them to construct the $N$-particle rotor-router aggregate
for $N=10^{10}$,
which would be far beyond the reach of a straightforward approach.
The pictures at 
\href{http://rotor-router.mpi-inf.mpg.de/}{{\tt http://rotor-router.mpi-inf.mpg.de/}}
show, for various choices of the design-parameters
and for various large values of $N$, what the aggregate looks like
if one starts with all rotors in the same state
and adds $N$ particles to the blob.

\subsection{\label{sec:organize} Self-organization}

\begin{figure}
\begin{center}
\scalebox{.5}{
\includegraphics{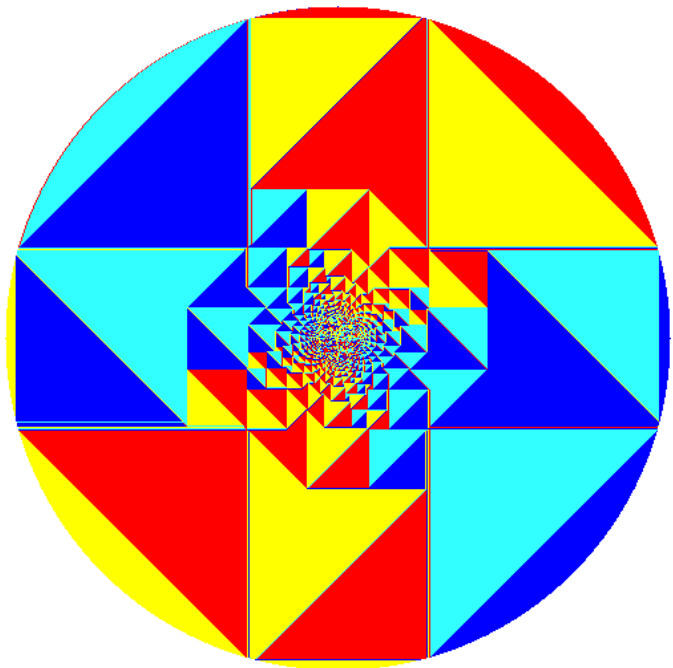}
}
\end{center}
\caption{\label{fig:vortex} Self-organization of rotor-routers in a disk.}
\end{figure}

We will not dwell on the self-organized criticality feature
of rotor-router systems,
though it was an essential part of the vision 
that led Priezzhev, Dhar {\it et al.\/} (1996) to introduce
the Eulerian walkers model in the first place
(see subsection~\ref{sec:sandpile}).
However, we will remark that
an important (though still poorly understood) feature 
of the pictures at Friedrich's website
is that some remarkably intricate and stable forms of order
are brought into existence by the rotor-router rule.
Figure~\ref{fig:vortex} shows
another instance of this sort of self-organization.
Here the underlying graph is the subgraph of the square grid $\Z \times \Z$
consisting of all vertices $(i,j)$
with $i^2 + j^2 \leq (250)^2$ (a discrete disk)
along with all the neighbors of those vertices
that do not belong to the disk itself (a discrete corona);
vertices in the corona correspond to output lines,
and there is an input line to $(0,0)$.
This corresponds to an electrical network
in which the center of the disk is clamped to voltage 0
and vertices in the corona are clamped to voltage 1.
The rotors are initially all pointing in the same direction.
The Figure shows the state of the rotors
after 1000 chips have passed through the system,
with the four colors corresponding to
the four states of the rotors.

\section{\label{sec:other} Other models}

\subsection{\label{sec:sandpile} Sandpile model aka chip-firing}

The 2-state rotors we have discussed so far
alternate between sending a chip along one arc
and sending a chip along the other.
A different approach to derandomization
is the sandpile model, or chip-firing model,
where the processor at a vertex alternates between holding a chip at $v$
and sending a chip simultaneously to both neighbors of $v$.
That is, if a vertex has no chips, and a chip arrives,
it must wait there until a second chip arrives,
at which moment the two chips can leave the vertex,
with one chip leaving along each of the two arcs.
(Since the chips are indistinguishable,
we need not to worry about
deciding which chip travels along which arc.) 
More generally, if a vertex with $n$ outgoing arcs 
is occupied by $\geq n$ chips,
we may send 1 chip along each arc,
but we are not permitted to move any of the chips at $v$
if there are fewer than $n$ of them.

Most of what has been said above about rotor-routing
applies as well to chip-firing (see Engel, 1975 
and 1976), 
including the constant/$N$ bound on discrepancy,
although the constant here tends to be larger.
For a discussion of relationships
between chip-firing and rotor-routing,
see Kleber (2005) 
and Holroyd {\it et al.\/} (2008).  

The sandpile model was invented 
by Bak, Tang, and Wiesenfeld (1987), 
and most the early rigorous theoretical analysis of the model
is due to Deepak Dhar (see e.g., Dhar 1999). 
Dhar and collaborators also explored the rotor-router model
under the name of the ``Eulerian walkers model''
(Priezzhev {\it et al.}, 1996,  
Shcherbakov {\it et al.}, 1996, 
and Povolotsky {\it et al.}, 1998). 
Dhar (1999) 
proposed that both the rotor-router model and sandpile model
can be viewed as special cases of a more general
``abelian distributed processors model''.
This is related to the observation
that networks of rotor-routers
themselves behave like rotor-routers.
For instance, the binary counter of Figure~\ref{fig:counter}
acts like a $2^m$-state rotor,
while the resistive network simulator of Figure~\ref{fig:electric}
acts like a 20-state rotor
(this is the order of the element associated with the vertex $a$
in the sandpile group associated with the graph;
see Holroyd {\it et al.}, 2008 
for a discussion
of the relation between rotor-routing and the sandpile group).
More generally, given any network of rotor-routers,
if one looks at a connected sub-network of rotors,
one obtains a multi-input, multi-output finite-state machine
that has the abelian property:
if one hooks such sub-networks together
and passes chips through the compound network,
the order in which the sub-networks
process the chips passing through them
does not affect the final outcome.

\subsection{\label{sec:cs} Synchronous network flow}

Cooper and Spencer (2006) 
studied rotor-routing in a slightly different setting,
where we have a (finite) number of chips initially placed in a graph
and we advance each of them $t$ steps in tandem
(every chip takes a first step,
then every chip takes a second step,
etc., for $t$ rounds).
Note that when we move chips in tandem in this way,
the abelian property does not apply;
for instance, we cannot move one chip $t$ steps,
then move another chip $t$ steps, etc.,
and be assured of reaching the same final state.
For a more detailed discussion of this point, 
see Figure 8 of Holroyd {\it et al.\/} (2008) 
and the surrounding text.

Cooper and Spencer showed that when the graph is $\Z^d$,
and the initial distribution of chips is restricted
to the set of vertices whose coordinates have sum divisible by 2,
then the discrepancy between the number of chips
at location $v$ at time $t$ under rotor-routing
and the expected number of chips
that would be at $v$ at time $t$ under random routing
is at most a constant that depends on $d$ ---
not on $v$, $t$, the distribution of the chips at time 0,
or the initial configuration of the rotors.
This result assumes that all rotors turn counterclockwise
(or equivalently, all rotors turn clockwise).
Using a different sort of 4-state rotor at each vertex
merely changes the constant.
(Of course we are assuming that the four states of the rotor at $v$
send a chip to the four neighbors of $v$.)
Similar results hold in higher-dimensional grids,
though the constants are bigger.
For articles that pursue this further,
see Cooper {\it et al.\/} (2006), 
Cooper {\it et al.\/} (2007), 
and Doerr and Friedrich (2009). 

\subsection{\label{sec:grow} Growth model}


Imagine that the sites of $\Z^2$ start out being unoccupied,
and that we use random walk or rotor-walk
to fill in the vicinity of $(0,0)$ with a growing ``blob''.
Specifically, we release a particle from $(0,0)$
and let it walk until it hits a site
that is not yet part of the blob;
then this site is added to the blob
and the particle is returned to $(0,0)$
to start its next walk.
In the case where the walk is a random walk,
this is the Internal Diffusion-Limited Aggregation Model,
invented independently by physicists (Meakin and Deutch, 1986) 
and mathematicians (Diaconis and Fulton, 1991); 
results of Lawler, Bramson, and Griffeath (1992) 
show that with probability 1, 
the $N$-particle blob, rescaled by $\sqrt{N/\pi}$,
converges to a disk of radius 1.
In the case where the walk is a rotor-walk,
with clockwise or counterclockwise progression of the rotors,
and with all rotors initially aligned with one another,
this is the ``rotor-router blob''
introduced by this author circa 2000
and analyzed by Levine and Peres (2009). 
Whereas the $N$-particle IDLA blob appears to have
radial deviations from circularity on the order of $\log N$,
the deviations from circularity for the $N$-particle rotor-router blob
appear to be significantly smaller; 
see Friedrich and Levine (2010).  
In particular, it is possible that
the deviation, as measured by
the difference between the circumradius and inradius of the blob,
remains bounded for all $N$.
Here one should measure the circumradius and inradius
from $(1/2,1/2)$ rather than $(0,0)$,
since there is both theoretical and empirical evidence
for the conjecture that the center of mass of the blob
approaches $(1/2,1/2)$.

In any case, it appears that,
using completely local operations, the network can ``tell''
which of two far-away points is closer to $(1/2,1/2)$
as long as $|r_1-r_2|$ is not too small compared to $r_1+r_2$,
where $r_1$ and $r_2$ are the points'
actual distances from $(1/2,1/2)$.
There is thus a sense in which rotor-router aggregation
computes approximate circles.
However, it should be mentioned that
this computation is not as robust as the rotor-router approach
to estimating escape probabilities and effective conductances.
For instance, simply by changing
the behavior of the rotors on the coordinate axes,
one can dramatically change the shape of the blob;
see Kager and Levine (2010). 

It has also been shown that rotor-routers provide a good approximation 
not just to internal DLA with a single point-source
but also to more general forms of internal DLA,
describable as PDE free-boundary problems;
see Levine and Peres (2007). 

\section{\label{sec:conclude} Conclusions}

Although we have focused above on computing effective conductances,
rotor-routing networks also measure voltages and currents.
For instance, to measure the current flow
in a resistive circuit between two neighboring nodes $v$ and $w$,
we need only look at the net flow of chips from $v$ to $w$
(that is, the flow of chips from $v$ to $w$
minus the flow of chips from $w$ to $v$)
in the associated rotor-router network,
divided by the number of chips that have passed through the network.

The rotor-router model is nonlinear,
but because it approximates the divisible flow model,
it is in many respects exactly solvable.
In particular, there is an asymptotic sense
in which, as the number of indivisible particles 
that flow through a network goes to infinity,
the behavior of the system approaches
the behavior of the divisible model.
The same is true for the random-router model,
but the discrepancies go to zero more slowly.
Furthermore, the performance guarantees for rotor-routing are deterministic,
whereas the performance guarantees for random routing are random
(there is a small but non-zero probability
that the discrepancy will be much larger
than the $O(1/\sqrt{n})$ average-case bound).

A key property of the rotor-router model
is the existence of conserved bulk quantities
expressible as sums of locally-defined quantities.
In the examples we have studied here,
there is essentially only one such quantity
(the value of the chips-and-rotors configuration),
but the space of such invariants can be higher-dimensional.
Specifically, if one is solving a Dirichlet problem
where the value of a function is constrained at $m$ points
(as in the case of the system shown in Figure~\ref{fig:vortex}),
the space of linear dynamical invariants is $m$-dimensional.

On the other hand, the usual concept of dynamics-in-time
is in a sense inappropriate for this kind of system
(leaving aside the Cooper-Spencer sort of scenario),
since, when there are multiple chips in the system,
it is not meaningful to ask which chip should take a step next;
the number of chips that exit the network along a particular output line
is independent of the order in which steps are taken,
and indeed there can be two events in the network
for which it does not make sense to ask which one occurs first, 
since the time-order of the events depends on the choice of dynamical path.
Our notion of dynamics should be flexible enough
to accommodate this symmetry.
Note that in some applications
one can choose which event will occur next
according to some probability distribution,
and then the system becomes a Markov chain
with stochastic dynamics in ordinary time,
but imposing such a probability law
is extrinsic to the system as we have described it above.
Rather, for systems like rotor-routing,
the dynamics is expressed
not in a {\it function\/} (given the current state of the system, 
here is what the next state of the system {\it must\/} be)
but in a {\it relation\/} (given the current state of the system, 
here is what the next state of the system {\it can\/} be).

Another noteworthy feature of rotor-routing
is the way networks store information.
In one sense, the information resides in the pattern of connections;
in another sense, information resides in the rotor-states,
and more specifically, in the numerical values of those states.

A consequence of the distributed way 
in which these networks store information
is their robustness in the presence of errors.
Rotor-router computations are not always robust
(the binary counter is not robust
under changes to its most significant bits,
and rotor-router aggregation has its own sort of sensitivity
to small perturbations; see Kager and Levine, 2010). 
But when the values of the rotor-states
are all significantly smaller than
the values on the network's output lines,
a small number of errors will not have a large effect
on the accuracy of the computation.
Part of the reason for this
is that the computation is intrinsic to the network's connectivity pattern
(as the behavior of random routing shows);
but the use of rotor-routing instead of random routing
reduces these errors even more.

Rotor-router computations have the feature
that, if you can correctly guess
the number of times each vertex emits a chip,
you can rigorously prove that your guess is correct
with much less work than is required
to derive the number of times each vertex emits a chip
by simulating the system.

Lastly, rotor-router computation
serves as an example of
{\it digital analogue computation}
(using here the root meaning of the term ``analogue'').
The concept of analogy is more crucial to the study of computation
than distinctions like discrete-versus-continuous
or even deterministic-versus-random.
Indeed,
as the three network routing models of section~\ref{sec:three} demonstrate,
a discrete model can be an analogue of a continuous model,
and a deterministic model can be an analogue of a stochastic model.

\noindent
\begin{quotation}
\noindent
Thanks to Deepak Dhar, Tobias Friedrich, Lionel Levine, Cris Moore,
and an anonymous referee for their help during the writing of this article.
\end{quotation}

\nocite{*}
\bibliographystyle{plain}
\bibliography{rotor-xxx}

\providecommand{\noopsort}[1]{}\providecommand{\singleletter}[1]{#1}%
\begin{thebibliography}{10}

\bibitem{BTW}
P.~Bak, C.~Tang, and K.~Wiesenfeld.
\newblock Self-organized criticality: an explanation of the 1/f noise.
\newblock {\em Physical Review Letters}, 59(4):381--384, 1987.

\bibitem{CDST06}
J.~Cooper, B.~Doerr, J.~Spencer, and G.~Tardos.
\newblock Deterministic random walks.
\newblock In {\em Proceedings of the Workshop on Analytic Algorithmics and
  Combinatorics}, pages 185--197, 2006.

\bibitem{CDST07}
J.~Cooper, B.~Doerr, J.~Spencer, and G.~Tardos.
\newblock Deterministic random walks on the integers.
\newblock {\em European Journal of Combinatorics}, 28(8):2072--2090, 2007.

\bibitem{CS}
J.~Cooper and J.~Spencer.
\newblock Simulating a random walk with constant error.
\newblock {\em Combinatorics, Probability and Computing}, 15:815--822, 2006.

\bibitem{D}
D.~Dhar.
\newblock Studying self-organized criticality with exactly solved models.
\newblock arXiv:cond-mat/9909009, 1999.

\bibitem{DF}
P.~Diaconis and W.~Fulton.
\newblock A growth model, a game, an algebra, lagrange inversion, and
  characteristic classes.
\newblock {\em Rend. Sem. Mat. Univ. Politec. Torino}, 49(1):95--119, 1991.

\bibitem{DF09}
B.~Doerr and T.~Friedrich.
\newblock Deterministic random walks on the two-dimensional grid.
\newblock {\em Combinatorics, Probability and Computing}, 18:123--144, 2009.

\bibitem{DS}
P.~G. Doyle and J.~L. Snell.
\newblock {\em Random Walks and Electrical Networks}.
\newblock The Mathematical Association of America, 1984.
\newblock Revised version at arXiv:math/0001057.

\bibitem{E1}
A.~Engel.
\newblock The probabilistic abacus.
\newblock {\em Ed. Stud. Math.}, 6:1--22, 1975.

\bibitem{E2}
A.~Engel.
\newblock Why does the probabilistic abacus work?
\newblock {\em Ed. Stud. Math.}, 7:59--69, 1976.

\bibitem{FL}
T.~Friedrich and L.~Levine.
\newblock Fast simulation of large-scale growth models.
\newblock arXiv:1006.1003, 2010.

\bibitem{HLMPPW}
A.~E. Holroyd, L.~Levine, K.~M\'esz\'aros, Y.~Peres, J.~Propp, and D.~B.
  Wilson.
\newblock Chip-firing and rotor-routing on directed graphs.
\newblock {\em In and Out of Equilibrium 2, {\rm in ``Progress in
  Probability''}}, 60:331--364, 2008.
\newblock arXiv:0801.3306.

\bibitem{HP}
A.~E. Holroyd and J.~Propp.
\newblock Rotor walks and {M}arkov chains.
\newblock {\em Algorithmic Probability and Combinatorics, {\rm in
  ``Contemporary Mathematics''}}, 520:105--126, 2010.
\newblock arXiv:0904.4507.

\bibitem{KL}
W.~Kager and L.~Levine.
\newblock Rotor-router aggregation on the layered square lattice.
\newblock arXiv:1003.4017, 2010.

\bibitem{K}
M.~Kleber.
\newblock Goldbug variations.
\newblock {\em Mathematical Intelligencer}, 27(1):55--63, 2005.

\bibitem{LBG}
G.~F. Lawler, M.~Bramson, and D.~Griffeath.
\newblock Internal diffusion limited aggregation.
\newblock {\em Annals of Probability}, 20(4):2117--2140, 1992.

\bibitem{LP07}
L.~Levine and Y.~Peres.
\newblock Scaling limits for internal aggregation models with multiple sources.
\newblock {\em Journal d'Analyse Math\'ematique}.
\newblock To appear; arXiv:0904.4507.

\bibitem{LP09}
L.~Levine and Y.~Peres.
\newblock Strong spherical asymptotics for rotor-router aggregation and the
  divisible sandpile.
\newblock {\em Potential Analysis}, 30:1--27, 2009.

\bibitem{L}
Lionel Levine, 2009.
\newblock Unpublished memorandum.

\bibitem{LyP}
R.~Lyons and Y.~Peres.
\newblock {\em Probability on Trees and Networks}.
\newblock Cambridge University Press, 2010.
\newblock In preparation.

\bibitem{MD}
P.~Meakin and J.~M. Deutch.
\newblock The formation of surfaces by diffusion-limited annihilation.
\newblock {\em Journal of Chemical Physics}, 85:2320--2325, 1986.

\bibitem{MM}
C.~Moore and J.~Machta.
\newblock Internal diffusion-limited aggregation: parallel algorithms and
  complexity.
\newblock {\em Journal of Statistical Physics}, 99(3--4):661--690, 2000.

\bibitem{PPS}
A.M. Povolotsky, V.B. Priezzhev, and R.R. Shcherbakov.
\newblock Dynamics of eulerian walkers.
\newblock {\em Physical Review E}, 58:5449--5454, 1998.

\bibitem{PDDK}
V.~B. Priezzhev, D.~Dhar, A.~Dhar, and S.~Krishnamurthy.
\newblock Eulerian walkers as a model of self-organised criticality.
\newblock arXiv:cond-mat/9611019, 1996.

\bibitem{SPP}
R.R. Shcherbakov, V.V. Papoyan, and A.M. Povolotsky.
\newblock Critical dynamics of self-organizing eulerian walkers.
\newblock arXiv:cond-mat/9609006, 1996.

\bibitem{S}
F.~Spitzer.
\newblock {\em Principles of Random Walk}.
\newblock Springer-Verlag, 1976.

\end{thebibliography}

\end{document}